# Architecture of Text Mining Application in Analyzing Public Sentiments of West Java Governor Election using Naive Bayes Classification

Suryanto Nugroho
Master of Informatics Engineering,
Amikom Yogyakarta University
Jl. Ring Road Utara, Condong Catur, Sleman
Yogyakarta, Indonesia

Prihandoko
Department of Computer Science and Information Technology, Gunadarma University
Jl. Margonda Raya No.100, Pd. Cina, Beji, Depok, West Java, Indonesia

## ABSTRACT
The selection of West Java governor is one event that seizes the attention of the public is no exception to social media users. Public opinion on a prospective regional leader can help predict electability and tendency of voters. Data that can be used by the opinion mining process can be obtained from Twitter. Because the data is very varied form and very unstructured, it must be managed and uninformed using data pre-processing techniques into semi-structured data. This semi-structured information is followed by a classification stage to categorize the opinion into negative or positive opinions. The research methodology uses a literature study where the research will examine previous research on a similar topic. The purpose of this study is to find the right architecture to develop it into the application of twitter opinion mining to know public sentiments toward the election of the governor of west java. The result of this research is that Twitter opinion mining is part of text mining where opinions in Twitter if they want to be classified, must go through the preprocessing text stage first. The preprocessing step required from twitter data is cleansing, case folding, POS Tagging and stemming. The resulting text mining architecture is an architecture that can be used for text mining research with different topics.

## 1. INTRODUCTION
Online user growth gives a direct influence on the use of social media. The Association of Internet Service Providers Indonesia (APJII) said the number of internet users in Indonesia reached 132.7 million people based on a survey conducted in 2016 this means an increase of 51.8% compared to the end of 2014 which reached 88.1 million. The survey data suggests that there are three main reasons Indonesians use the internet. The three goals are to access social/communication facilities, daily news sources, and keep up with the times.

Social media in this era of globalization is ubiquitous and widely used for the benefit of society. In its implementation, social media is more widely used for buying and selling activities, conveying information, even as a medium to express themselves. Twitter is a social network that is often used as a means of communication, media for promotion, and political campaigns. Twitter is a social network that has unique characteristics and shapes with particular symbols or regulations. Twitter users can only send and read blog messages as in general with an optimal limit of 140 characters, the word is known as tweets [1]. Each tweet in the post users of various kinds according to the will of the user. They can be opinions, suggestions, or criticism around specific topics. The multiple postings, as well as the abundant usage of non-standard language on a tweet, is the reason for the need for sentiment analysis.

The analysis of sentiments or so-called opinion mining aims to analyze, understand, process, and execute textual data in the form of opinions on entities such as organizations and specific topics to obtain information [2]. Meanwhile, according to [3] analysis of sentiment in a sentence describes the consideration of the assessment of specific entities or events. The sentiment analysis also focuses on the processing of opinions that contain polarity, in other words, has a positive or negative sentiment points. The problems in sentiment analysis are difficult to define, define the concept of the issues, sub-problems, and objectives that serve as a performance framework in various studies [2].

Text mining refers to the process of retrieving high-quality information from text. High-quality information is usually obtained through forecasting patterns and trends through means such as statistical learning patterns. Typical text mining processes include text categorization, text clustering, concept extraction/entity, the production of granular taxonomy, sentiment analysis, document conclusion, and entity relationship modeling (i.e., learning relationships between named entities)[4].

Extraction of information from unstructured text [5], which will help all types of users. Information retrieval is usually done in the form of pipeline network analysis. The pipe stage is formant conversion, sentence solvers, tokenization, word stemming and token annotations. Temporal data explain mining techniques in decision support systems[6]. In the relationship between the events affecting the decision are discussed. This link is defined using unattended data mining learning techniques. It not only describes the connections between events, but it also extracts the new patterns and boundaries that exist within the system. Unattended learning and learning techniques supervise these two methods of mining engineering. Controlled learning techniques are used in predictive statistical methods whereas unsupervised learning methods do not apply dependent variables. It looks for patterns and events[7].

The election of governors and representatives of West Java period 2018-2023 will be held in 2018, so that live a few more months. In the pre-implementation and implementation of West Java governor election, there are various opinions and responses with positive and negative sentiments on Twitter.

The problems that arise when analyzing all sentiment results and classifying tweets on Twitter manually then it will take time and effort that many and not accurate. In the previous period, predictions of election results were conducted by survey agencies and the quick count. Sentiment analysis used in social media to election the of the governor of west java become one alternative with an inaccurate quick count which has been done by survey institute. The tendency of users in expressing opinions through social media can be a way of knowing electability a candidate for governor in an election event.

In doing text mining and sentiment analysis is also needed an application architecture so that what is done following the path that has been determined, so that get maximum results. This paper aims to provide a recommendation of the application architecture in text mining or opinion mining concerning the selection of West Java governor on social media twitter Indonesia by using Naive Bayes classification.

## 2. BASIC THEORY

Sentiment analysis or commonly called opinion mining is one branch of Text Mining research. Opinion mining is computational research of opinions, sentiments, and emotions that are textually expressed. If given a set of text documents containing opinions about an object, then opinion mining aims to extract the attributes and components of the commented object on each paper and to determine whether the comment is positive or negative [8]. Its data source can distinguish sentiment Analysis, some of the most commonly used levels of Sentiment Analysis are Sentiment Analysis at the document level and Sentiment Analysis at sentence level [9]. Based on the level of data source Sentiment Analysis is divided into two major groups namely [9]: Coarse-grained Sentiment Analysis and fined-grained Sentiment Analysis

In Coarse-grained Sentiment Analysis, Sentiment Analysis is done at the document level. Broadly speaking the primary focus of this type of Sentiment Analysis is to assume the entire contents of the document as a positive sentiment or negative sentiment. Fine-grained Sentiment Analysis is Sentiment Analysis at the sentence level. The primary focus of the fined-grained Sentiment Analysis is to determine the sense in every sentence.

Text mining can be broadly defined as an intensive knowledge process in which users interact with document collections over time using a single analysis tool. Text mining attempts to extract useful information from data sources through the identification and exploration of unusual patterns. Text mining tends to lead to the field of data mining research. Therefore, it is not surprising that text mining and data mining are at the same architectural level [10]. Text mining may be regarded as a two-stage process beginning with the application of structures to text data sources and followed by the extraction of relevant information and knowledge from this structured text data using the same techniques and tools like data mining [10].

Text Preprocessing is the stage of the initial process of text to prepare the version into data to be processed further. An existing text should be separated; it can be done on several different levels. A document or tweet can be broken down into chapters, sub-chapters, paragraphs, sentences and ultimately into pieces of words/tokens. Also at this stage, the existence of digit numbers, capital letters, or other characters is removed and changed [10].

The Naive Bayes classifier algorithm is an algorithm used to find the highest probability value for classifying test data in the most appropriate category [10]. In simple terms, an NBC assumes that the presence (or absence) of specific features of a class is unrelated to the presence (or absence) of other features. For example, the fruit will be considered an apple if it is red, round, and about 4 inches in diameter. An NBC assumes that all traits contribute independently to the probability that the fruit is an apple. An advantage of NBC is that it requires a small amount of training data to estimate the parameters (average and variant of the variables) needed for classification. Since variables are assumed to be independent, only variants of the variables for each class need to be determined and not the overall covariance matrix[10].

## 3. RESEARCH METHODOLOGY

The study was conducted with several activities to analyze the processes needed to process opinion data from Twitter to the conclusion of the percentage of positive and negative opinions for a brand. The research steps are as follows.

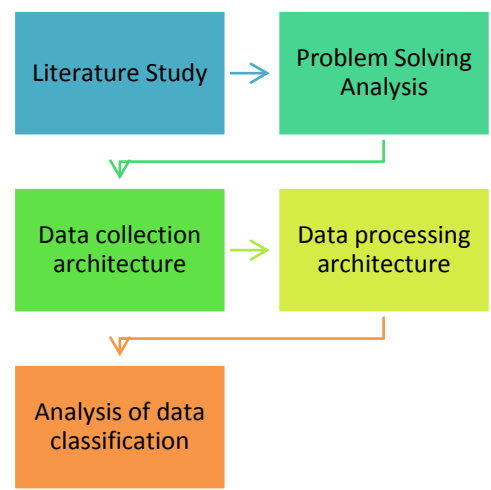

**Fig 1: Research Methodology**

From Figure 1, we can see the overall process in the implementation of this research method. There are five steps done in this research, among others:

a. Doing literature review
Research is continued by finding solutions to problems by studying similar research on opinion mining and text mining.

b. Problem-solving analysis
The problem is how to form unstructured opinions into semi-structured then classify them in positive opinions and negative opinions.

c. Data collection architecture
Designing an architecture used for tweet data collection from Twitter to be processed.

d. Data processing architecture
Designing an architecture used for processing tweet data that has been collected, so that data is ready to be classified.

e. Analysis of data classification
It is a step taken to classify data already in process to generate positive and negative sentiments.

## 4. DISCUSSION
### 4.1 Research Approach
The type of research used in this study is qualitative, i.e., research conducted by collecting data in the form of words or sentences from the source information that the author made the object, in this case, twitter.

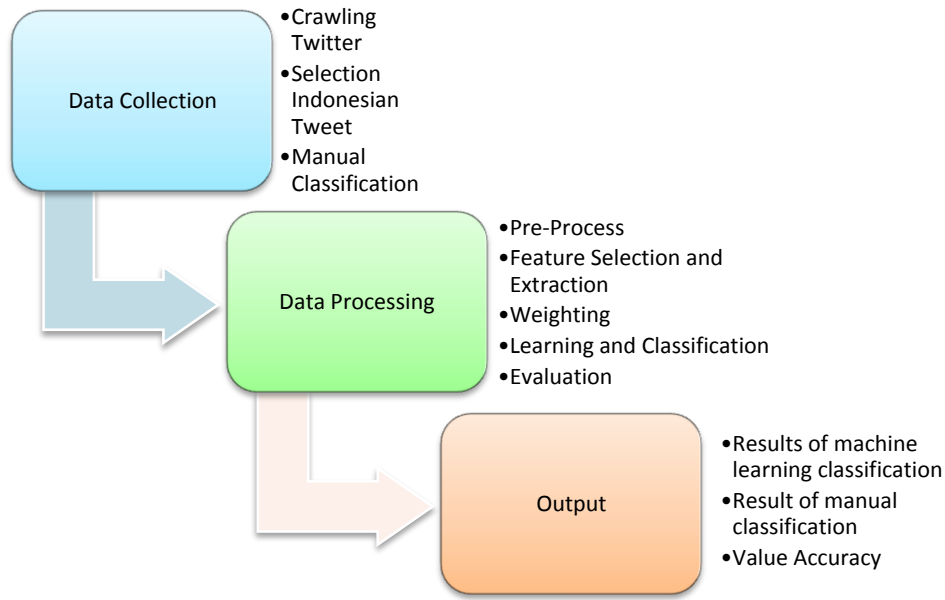

**Fig 2: Step Research**

From Figure 2, we can see the overall process in the implementation of this research method. Where the data collected first is a tweet with a unique keyword associated with public figures who will go forward as governor or deputy governor of western Java with a tweet in the Indonesian language that has been determined from the Twitter account Indonesia. Then the use of software to support the classification method and the last is the output in the form of data analysis, the value of accuracy and classification results.

In Figure 2 shows the step of the analysis of sentiment analysis on the selection of governors in the province of western Java Indonesia. Three main steps are done, among others:

a. Data collection
This step includes some activities ranging from text mining on twitter or often called crawling data and tweets that have been in the selection made only tweet Indonesian language used after that done the classification manually which will be compared with the classification system.

b. Data Processing
Activity on processing include text preprocessing on tweet data, then performing selection and feature extraction after it conducts the weighting and the last one is an evaluation.

c. Output
The output produced by data processing is the accuracy of the classification done in the system compared to the manual classification. The algorithm that can be used is Naive Bayes.

### 4.2 Data Collection Methods
Before conducting sentiment analysis in the Indonesian language, the initial process which must be done in this research is collecting corpus. Data used for the collection of corpus obtained from Twitter. The method used in collecting corpus data refers to a previous study of "Twitter as a Corpus for Sentiment Analysis and Opinion Mining" [11], plus another step at the end of the process to provide Twitter data using the Indonesian language. In the study mentioned that to separate the positive and negative sentiments of Twitter can be done by searching by using keywords in the form of emotion icons.

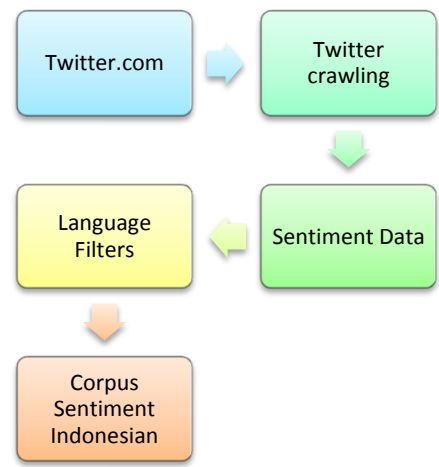

**Fig 3: Architecture Data Collection Methods**

In Figure 3 the process of collecting tweet data from crawling tweet data to generate sentiment data, after which will be filtering the language to be taken is a tweet that uses the Indonesian expression, and the result of filtering will get corpus Indonesian language. In the crawling stage of data tweet using hashtags #pilgubJabar, #ridwankamil, #deddymizwar, #dedimulyadi, #pilkadajabar. Crawling data tweet will also take data from each candidate governor and vice-governor of western Java who will fight in elections in 2018.

The process of crawling on Twitter is done by utilizing the facility application interface (API) that has been provided.

The maximum character length of a single sentence tweet is 140 characters, so one data tweet is a sentence representing a sentiment. Emotion icons used as keywords are ":)" and ":(", use these emotion icons according to the twitter search guidelines, where to look for tweets that have a positive expression using the ":)" emotion icon and to search for tweets that have negative expressions using the ":(" emotion icon.

## 4.3 Data Analysis Methods

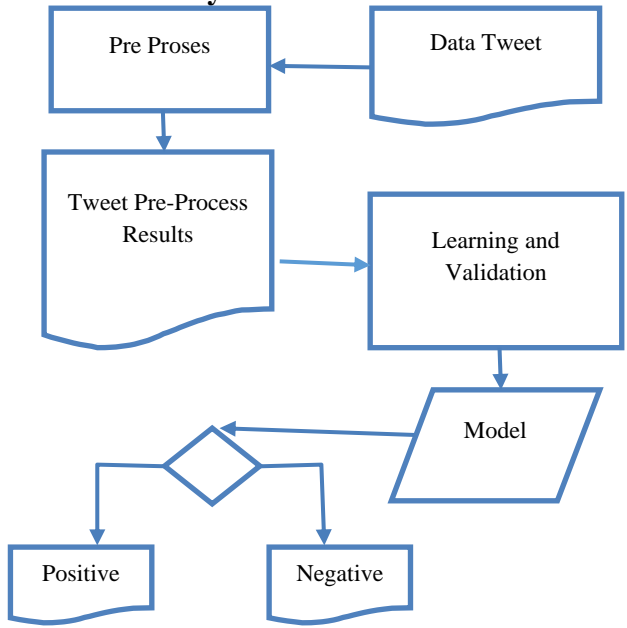

**Fig 4: The Flow of Data Processing**

Figure 4 shows how the data processing flow of sentimental research analyzes the selection of West Java governor. The data required for this classification process consists of learning data and validation data. Before the learning process is done, pre-processed on the data corpus (tweet).

Pre-Process

The pre-process needs to be done before the classification process so that the dimension of the vector space model becomes lower than that of the pre-process. At least by creating a low-dimensional vector space model, then the classification process will grow faster.

The purpose of this pre-processed tweet is:

1. Eliminate unscreened tweets

2. Uniform the form of words

3. Reduce word volume

The steps performed on the pre-process tweets can be seen in Figure 5.

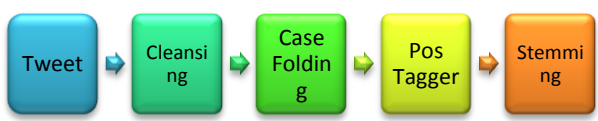

**Fig 5: Architecture Pre-Process Data Tweet**

Cleansing
The cleansing process is the process of cleaning tweets from unnecessary words to reduce noise in the classification process.

Stopwords
Stopwords are common words that usually have little effect in a text, such as "and," "but," etc. The words contained in the stopwords list are omitted.

Case Folding
Case folding is the process of converting all the letters in the tweet into lowercase (the letter 'a' up to the letter 'z'). Characters other than letters will be removed because they are considered as delimiters.

Feature Selection
The most important thing in the process of sentiment classification is the selection of features used for classification. Some commonly used standard features are unigram, each word in the tweet will be considered a feature (in some previous research called POS Tagger). These functions and values will then be stored in a vector for classification purposes.

Part of Speech (POS) Tagger
Tagger POS is a process of giving classes to a word. According to Dr. Goris Keraf (1979), almost all the grammars that exist today divide words based on Aristotle.

Stemming
The process of stemming is the process of returning the words that have been processed in the previous stage to be the primary word. This process will reduce the variations of words that have the same variety.

## 4.4 Classification

In the Naïve Bayes classifier algorithm, each document is represented by the attribute pair "X1, X2, X3 ... Xn" where X1 is the first word, X2 is the second word and so on. While V is the set of Tweet categories. At the time of classification, the algorithm will search for the highest probability of all categories of documents tested (Vmap), where the equation is as follows:

$$V_{MAP} = \underset{V_j \, e \, V}{\arg\max} \frac{P(x_1, x_2, x_3, \dots x_n | V_j) P(V_j)}{P(x_1, x_2, x_3, \dots x_n)}$$

For P (X1, X2, X3 ... Xn) the value is constant for all categories (Vj) so that the above equation can be written

$$V_{MAP} = \underset{V_j \, e \, V}{\arg\max} P(x_1, x_2, \dots x_n | V_j) P(V_j)$$

The above equation can be simplified as follows:

$$V_{MAP} = \underset{V_j \, e \, V}{\arg\max} \prod_{i=1}^{n} P(x_i | V_j) P(V_j)$$

Information:
Vj = tweet category j = 1,2,3, .. n Where in this study

j1 = negative sentence tweets category,

j2 = positive tweet category sentiment,

j3 = neutral tweet sentiment category.

P (Xi | Vj) = Probalitas Xi in category Vj

P (Vj) = Probability of Vj

For P (Vj) and P (Xi | Vj) are calculated during training where the equation is as follows:

$$P(V_j) = \frac{|docs\ j|}{|contoh|}$$

$$P(x_i|V_j) = \frac{n_k + 1}{n + |kosakata|}$$

Information:

| docs j | = number of documents per category j

| contoh | = number of documents from all categories

nk = number of times the occurrence of each word

n = number of times word occurrence of each category

| kosakata | = number of all words from all categories.

## 5. CONCLUSION
Based on what has been described above, it can be concluded things as follows:

1. Based on the review literature that has been done to obtain knowledge that Twitter opinion mining is part of the text mining where opinions in twitter if you want to be classified must go through the preprocessing text stage first.
2. The preprocessing step required from twitter data is cleansing, case folding, POS Tagging and stemming
3. Architecture that formed is the flow of processing logic from unstructured twitter opinion data into positive or negative sentiments against the figure who advanced the election of candidates for governor of West Java.

## 6. ACKNOWLEDGMENTS
Thank you the author to say to the Amikom Yogyakarta University as an alma mater author who has given an opportunity to write this paper. The authors are also grateful to the Aptikom distance education program for the support of lecturers and academic departments so that this article has been completed.